\documentclass[onecolumn, draftclsnofoot, 12pt]{IEEEtran}
\usepackage{amsmath}
\usepackage{amssymb}
\usepackage{amsfonts}
\usepackage{algorithm}
\usepackage{dsfont}
\usepackage{graphicx}
\usepackage{epsfig}
\usepackage{subfigure}
\usepackage{psfrag}
\usepackage{xcolor}

\title{CoMP in the Sky: UAV Placement and Movement Optimization for Multi-User Communications}
\author{Liang Liu, Shuowen Zhang, and Rui Zhang
\thanks{The authors are with the Department of Electrical and
Computer Engineering, National University of Singapore, Singapore
(e-mails: \{eleliu,elezhsh,elezhang\}@nus.edu.sg).}}

\begin{document}
\maketitle \thispagestyle{empty} \vspace{-0.3in}

%
%

\newtheorem{definition}{Definition}
\newtheorem{assumption}{Assumption}
\newtheorem{lemma}{\underline{Lemma}}
\newtheorem{example}{Example}
\newtheorem{theorem}{Theorem}
\newtheorem{proposition}{Proposition}
\newtheorem{conjecture}{Conjecture}
\newtheorem{remark}{Remark}
\newcommand{\mv}[1]{\mbox{\boldmath{$ #1 $}}}


\begin{abstract}
Driven by the recent advancement in unmanned aerial vehicle (UAV) technology, this paper proposes a new wireless network architecture of \emph{coordinate multipoint (CoMP) in the sky} to harness both the benefits of interference mitigation via CoMP and high mobility of UAVs. Specifically, we consider uplink communications in a multi-UAV enabled multi-user system, where each UAV forwards its received signals from all ground users to a central processor (CP) for joint decoding. Moreover, we consider the case where the users may move on the ground, thus the UAVs need to adjust their locations in accordance with the user locations over time to maximize the network throughput. Utilizing random matrix theory, we first characterize in closed-form a set of approximated upper and lower bounds of the user's achievable rate in each time episode under a realistic line-of-sight (LoS) channel model with random phase, which are shown very tight both analytically and numerically. UAV placement and movement over different episodes are then optimized based on the derived bounds to maximize the minimum of user average achievable rates over all episodes for both cases of full information (of current and future episodes) and current information on the user's movement. Interestingly, it is shown that the optimized location of each UAV at any particular episode is the weighted average of the ground user locations at the current episode as well as its own location at the previous and/or next episode. Finally, simulation results are provided to validate and compare the performance of the proposed UAV placement and movement designs under different practical application scenarios.
\end{abstract}


\section{Introduction}\label{sec:Introduction}
Leveraging cooperative communications over multiple transmission/reception points, coordinated multipoint (CoMP) \cite{Irmer11} has been incorporated into the current LTE (long-term evolution)-Advanced Releases \cite{3GPP} as an effective technique for mitigating inter-cell interference and exploiting the benefits of the currently deployed distributed antenna systems (DAS) \cite{Kerpez96}. For the fifth-generation (5G) cellular networks on the roadmap, another variant of CoMP, namely, cloud radio access network (C-RAN) \cite{Dittmann15,Simeone16}, is envisioned to be a promising candidate to harness the gain of cloud computing and achieve up to 1000 times of throughput improvement over today's fourth-generation (4G) cellular networks. For both DAS and C-RAN architectures, a proper deployment of the remote antenna units (RAUs) or remote radio heads (RRHs) is crucial to achieve superior channel qualities for all users and thus maximize the network throughput. With various design objectives, the RAUs/RRHs deployment problems have been extensively studied in the literature (see e.g., \cite{Wang09,Goldsmith11,Park12}). However, in practice, the users are moving over time, making a fixed deployment of RAUs or RRHs difficult to provide continuously high-quality services to the mobiles.

On the other hand, unmanned aerial vehicles (UAVs) have recently found wide applications in wireless communication systems for coverage extension \cite{Buttazzo15,Mozaffari16,Lyu17,Yanikomeroglu16} and capacity enhancement \cite{UAVrui16,He17,Wu17}, driven by the pioneering efforts of Google (Project Loon) and Facebook (Project Aquila). Deployed as flying base stations (BSs) in the sky, UAVs are able to adjust their locations dynamically and swiftly to provide flexible and on-demand services to the ground users according to their real-time locations. Further, the altitude of the UAVs is usually in the range from tens of meters (m) to about 100 m, making the favorable line-of-sight (LoS) channel available for the ground users without deep fading \cite{Qualcomm}. These notably promissing features of the UAV-ground communications motivate this paper to investigate a new CoMP-based wireless system with UAV-mounted RAUs/RRHs in the sky.

\subsection{Prior Work}
Placement of RAUs/RRHs is an important design problem for optimizing the system performance of DAS/C-RAN. For the case of uplink communications, \cite{Wang09} studies the RAU placement design in a multi-user DAS system, and proposes a squared distance criterion for maximizing a lower bound of the cell averaged ergodic capacity. On the other hand, for the case of downlink communication, \cite{Goldsmith11} considers the RAU placement design in a single-user DAS system, and proposes an iterative algorithm for maximizing the user ergodic rate based on stochastic approximation method. Moreover, under the assumption of circular RAU layout, \cite{Park12} investigates the RAU placement design in multi-user DAS systems for the purpose of maximizing the expected signal-to-noise ratio (SNR) with simple transmit beamforming strategies, e.g., maximal ratio transmission (MRT). Although the initial objective of the above works is to optimize the user ergodic rate, which is one of the most relevant performance metrics for DAS (or C-RAN), they consider alternative performance metrics in the placement design problems due to the difficulty in characterizing the user ergodic rate, which is a complicated function of the RAU or RRH locations.

Different from DAS or C-RAN with static RAUs or RRHs, the fully controllable mobility of the UAV offers more design degrees of freedom for throughput enhancement \cite{UAVrui17}. Specifically, the communication distance between the UAV and ground users can be significantly shortened via proper UAV trajectory design, which essentially leads to increased channel power gain under the LoS channel model and thus further opportunities for user rate maximization. For example, \cite{He17} considers a single-UAV and multi-user setup and proposes a practical fly-hover-and-communicate protocol to serve the users. Furthermore, \cite{Wu17} studies the trajectory design problem under a multi-UAV and multi-user setup, where the UAV trajectory is jointly optimized with the transmit power control and user scheduling and association to maximize the minimum rate of all users in a given finite period.

Although the above works have demonstrated the effectiveness of UAV trajectory design in optimizing performance of UAV-enabled communications, there still exist challenging issues that are unaddressed yet. First, in \cite{He17,Wu17}, the ground users are assumed to be static, while in practice their locations may change over time. Additional consideration of the user movement will significantly increase the complexity of the UAV trajectory design. Next, for the multi-UAV scenario in \cite{Wu17}, inter-user interference is mitigated via trajectory design and power control only, without considering CoMP-based cooperation. With coordinated UAVs, linear beamforming can be performed across the signals transmitted/received by all the UAVs to exploit the co-channel cross-links to achieve additional beamforming/spatial multiplexing gains. However, in this case, not only the amplitude but also the phase of the UAV-user channel will affect the user achievable rate, since the beamforming gain heavily relies on the channel phase. This is in sharp contrast to the scenario with single UAV or multiple UAVs without signal-level coordination, in which user rates only depend on the channel gains.

\subsection{Main Contributions}
\begin{figure}[t]
  \centering
  \includegraphics[width=10cm]{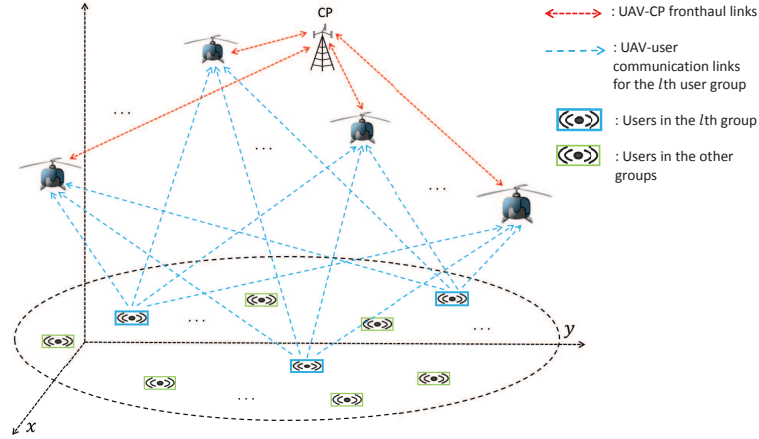}
  \vspace{-5mm}
  \caption{Schematic of a UAV-enabled CoMP system.}\label{System}
  \vspace{-10mm}
\end{figure}

In this paper, we consider uplink transmission in a UAV-enabled wireless communication system, where multiple single-antenna ground users send independent information to multiple single-antenna UAVs in the sky. We propose a new network architecture, namely, \emph{CoMP in the sky}, to leverage both the cooperation gain of CoMP for mitigating the inter-user interference and the high mobility of UAVs for providing strong channel gains to the moving ground users constantly. Specifically, instead of decoding the user messages locally, each UAV forwards its received signals from the ground users to a central processor (CP) via separate wireless fronthaul links (e.g., by using the millimeter wave (mmWave) band \cite{Dong16}), which  performs zero-forcing (ZF) beamforming across the signals from all the UAVs to jointly decode the user messages as illustrated in Fig. \ref{System}. Different from the static RAUs or RRHs in the conventional CoMP systems, the placement and movement of the UAVs are designed/updated over time in accordance with the user movement to achieve full coverage of the users and maximize the system average throughput over time. The main contributions of this paper are summarized as follows.
\begin{itemize}
\item First, we propose a practical transmission protocol for the above new network architecture. Specifically, the transmission time is divided into episodes, and the ground users are divided into groups. With sufficiently small episode, the user locations can be assumed to be static within each episode, but vary over different episodes. Within each episode, the users in the same group are served simultaneously in the same time/frequency block via spatial multiplexing, while users in different groups are served over orthogonal time/frequency blocks. We assume that each ground user has a delay constraint that it needs to be served within each episode.
\item Next, we propose a realistic \emph{LoS channel model with random phase} to characterize the user-UAV communication channel within each episode. Specifically, despite the user/UAV location variation, the amplitude of each UAV-user channel within one episode is assumed to be constant since the location variation over one episode (as long as the episode is sufficiently small) is practically negligible as compared to the altitude of the UAVs (say, 1 m versus 100 m). However, since channel phase is much more sensitive to the user/UAV location variation due to the small wavelength, it changes dramatically over different coherence intervals in each episode. Such a new model thus captures not only the essence of the traditional LoS channel -- constant amplitude, but also the unique feature of our considered UAV-ground channel -- fast phase variation.
\item Based on the above, this paper formulates a general problem to optimize the placement and movement of UAVs over a given number of time episodes so that the minimum achievable average rate of all the users over this time duration is maximized. In contrast to previous works with the focus on static ground users, the formulated problem applies to more challenging scenarios with moving users. In particular, three special cases of the formulated problem are considered to model three application scenarios, respectively: dynamic UAV placement with full user movement information (applicable to the case when the movement of the ground users (e.g., robots) is pre-programmed and known by the UAVs/CP ahead of time), dynamic UAV placement with current user movement information (applicable to the case when the movement of ground users is random and the UAVs merely know their locations at the current episode), and static UAV placement with full user movement information (applicable to the case when the UAVs need to hover at fixed locations to reduce the energy consumption of flying). Note that for the third scenario with static UAV placement, if we further assume a special case with fixed user locations, the corresponding results also apply to the conventional RAUs or RRHs deployment problem \cite{Wang09,Goldsmith11,Park12}.
\item By applying the random matrix theory, this paper derives tight and closed-form approximations of the user achievable rate over each episode under the considered LoS channel model with random phase. The obtained user rate approximations are then used to simplify the minimum-rate maximization problem for the UAV placement/movement design. By introducing auxiliary variables and applying the successive convex approximation technique, we are able to obtain locally optimal solutions to the approximated problems that satisfy all the Karush-Kuhn-Tucker (KKT) conditions. Further, by analyzing the KKT conditions of the problems, it is revealed that for the mobile UAV case, at each time episode, the optimized location of each UAV is a weighted average of user locations at the current episode as well as its own location at the previous and/or next episode, where the weights depend on the optimal dual variables of the considered problems.
\end{itemize}

\subsection{Organization}
The rest of this paper is organized as follows.
Section \ref{sec:System Model} describes the system model for our considered UAV-enabled CoMP.
Section \ref{sec:Problem Formulation} formulates various UAV placement and movement optimization problems to maximize the minimum of average user rates under different application scenarios.
Section \ref{sec:User Average Common Rate Characterization} presents approximated lower and upper bounds for the achievable minimum rate in closed-form.
Section \ref{sec:Proposed Algorithm} proposes efficient algorithms to solve the formulated problems.
Section \ref{sec:Numerical Examples} provides the numerical
simulation results to evaluate the performance of proposed algorithms. Finally, Section \ref{sec:Conclusion} concludes
the paper.

{\it Notation}: Scalars are denoted by lower-case letters, vectors
by bold-face lower-case letters, and matrices by
bold-face upper-case letters. $\mv{I}$ and $\mv{0}$  denote an
identity matrix and an all-zero matrix, respectively, with
appropriate dimensions. For a square matrix $\mv{S}$, $[\mv{S}]_{k,k}$ denotes its $k$th diagonal element. For a matrix
$\mv{M}$ of arbitrary size, $\mv{M}^{H}$ and ${\rm rank}(\mv{M})$ denote its
conjugate transpose and rank, respectively. $\mathbb{E}[\cdot]$ denotes the statistical expectation. The
distribution of a circularly symmetric complex Gaussian (CSCG) random vector with mean $\mv{x}$ and
covariance matrix $\mv{\Sigma}$ is denoted by
$\mathcal{CN}(\mv{x},\mv{\Sigma})$; and $\sim$ stands for
``distributed as''. $\mathbb{C}^{x \times y}$ denotes the space of
$x\times y$ complex matrices. $\|\mv{x}\|$ denotes the Euclidean norm of a complex vector
$\mv{x}$.

\section{System Model}\label{sec:System Model}
Consider the uplink transmission in a UAV-enabled wireless communication system consisting of $M>1$ single-antenna UAVs and $\tilde{K}>1$ single-antenna ground users. As illustrated in Fig. \ref{System}, we assume that all UAVs are connected to one ground CP via high-speed fronthaul links, such that the UAVs serve as coordinated RAUs/RRHs in the sky to cooperatively receive information from the ground users. Specifically, we consider that the $\tilde{K}$ users are divided into $L\geq 1$ equal-sized groups, each consisting of $K=\frac{\tilde{K}}{L}$ users, with $K<M$.\footnote{We consider that $\frac{\tilde{K}}{L}$ is an integer for convenience.} We assume that all users in the same group are served simultaneously via space-division multiple access (SDMA), while different groups are served in orthogonal time and/or frequency dimensions.

We further assume that the communication from the ground users to the UAV network in any period of interest is divided into $N$ equal-duration episodes, as shown in Fig. 2. In each $n$th episode, the locations of the $k$th user of group $l$ and the $m$th UAV are denoted by  $(\tilde{a}_{k,l}[n],\tilde{b}_{k,l}[n],0)$ and $(\tilde{x}_m[n],\tilde{y}_m[n],H)$, respectively, in a three-dimensional (3D) Cartesian coordinate system, where $k=1,\cdots,K$, $l=1,\cdots,L$, $m=1,\cdots,M$, and $n=1,\cdots,N$. In the above, we have assumed that all UAVs are deployed at a fixed altitude of $H$ m, while our results can be generalized to the case with heterogeneous UAV altitudes. The distance between the $m$th UAV and the $k$th user of group $l$ at each $n$th episode is thus given by
\begin{align}\label{eqn:distance1}
\tilde{d}_{k,l,m}[n]=\sqrt{(\tilde{x}_m[n]-\tilde{a}_{k,l}[n])^2+(\tilde{y}_m[n]-\tilde{b}_{k,l}[n])^2+H^2},\quad \forall k,l,m,n.
\end{align}

Note that within each $n$th episode, the horizontal locations of each $k$th user of group $l$ and $m$th UAV are quasi-static since they may experience small random variations with respect to their nominal values denoted by $({a}_{k,l}[n],{b}_{k,l}[n])$ and $({x}_m[n],{y}_m[n])$, respectively. In other words, $\tilde{a}_{k,l}[n]$, $\tilde{b}_{k,l}[n]$, $\tilde{x}_m[n]$, and $\tilde{y}_m[n]$ can be modeled as random variables with mean values $a_{k,l}[n]$, $b_{k,l}[n]$, $x_m[n]$, and $y_m[n]$, respectively. Such location variation can be caused by slight user movement in each episode, e.g., audience movement around their seats in a stadium, or UAV antenna vibration resulted from atmospheric turbulence, which may be created by the rotary wing of the UAV. It is worth noting that compared to the altitude of UAVs, i.e., $H$, provided that the episode length is chosen sufficiently small with respect to the maximum UAV/user speed, the maximum horizontal displacement of the UAV/users can be assumed negligible such that the distance between the $m$th UAV and the $k$th user of group $l$ at each $n$th episode can be well-approximated by a fixed value determined by their nominal locations:
\begin{align}\label{eqn:distance}
\tilde{d}_{k,l,m}[n]\approx d_{k,l,m}[n]=\sqrt{({x}_m[n]-{a}_{k,l}[n])^2+({y}_m[n]-{b}_{k,l}[n])^2+H^2},\quad \forall k,l,m,n.
\end{align}

\subsection{Channel Model}
At each episode $n$, we assume that the equivalent complex baseband channel coefficient between each $m$th UAV and $k$th user, denoted by $h_{k,m}[n]$, follows an LoS model thanks to the high altitude of the UAVs, i.e.,
\begin{align}\label{LoS}
h_{k,m}[n]&=\sqrt{\frac{\tau_0}{\tilde{d}_{k,l,m}[n]^2}}e^{j\frac{2\pi}{\lambda}\tilde{d}_{k,l,m}[n]}=|h_{k,l,m}^{\mathrm{LoS}}[n]|e^{j\tilde{\theta}_{k,l,m}[n]}\\
&\approx \sqrt{\frac{\tau_0}{d_{k,l,m}[n]^2}}e^{j\tilde{\theta}_{k,l,m}[n]},\quad \forall k,l,m,n,\label{LoS2}
\end{align}
where $\tau_0$ denotes the channel power gain at the reference distance $d_0=1$ m; $\lambda=\frac{c}{f_c}$ denotes the wavelength, with $c$ denoting the speed of light and $f_c$ denoting the carrier frequency; $|h_{k,l,m}[n]|=\sqrt{\frac{\tau_0}{\tilde{d}_{k,l,m}[n]^2}}$ and $\tilde{\theta}_{k,l,m}[n]=\frac{2\pi}{\lambda}\tilde{d}_{k,l,m}[n]$ denote the amplitude and the phase of $h_{k,l,m}$, respectively.

Note that both $|h_{k,l,m}[n]|$ and $\tilde{\theta}_{k,l,m}[n]$ are dependent on $\tilde{d}_{k,l,m}[n]$ given in (\ref{eqn:distance1}), thus are generally time-varying within each $n$th episode. By noting that $\tilde{d}_{k,l,m}[n]\approx d_{k,l,m}[n]$ holds as shown in (\ref{eqn:distance}), $|h_{k,l,m}[n]|$ can be well-approximated as $|h_{k,l,m}[n]|\approx\sqrt{\frac{\tau_0}{d_{k,l,m}[n]^2}}$ as shown in (\ref{LoS2}), which is fixed within each $n$th episode. On the other hand, $\tilde{\theta}_{k,l,m}[n]$ is much more critically dependent on $\tilde{d}_{k,l,m}[n]$ compared to $|h_{k,l,m}[n]|$ and thus cannot be approximated as a constant of $\frac{2\pi}{\lambda}d_{k,l,m}[n]$. This is because $\lambda$ generally takes a very small value in practice, thus a slight deviation of $\tilde{d}_{k,l,m}[n]$ from $d_{k,m}[n]$ due to user and/or UAV location variation within the $n$th episode will have a significant impact on $\tilde{\theta}_{k,l,m}[n]$. To characterize the effect of user/UAV location variation on $\tilde{\theta}_{k,l,m}[n]$, we model $\tilde{\theta}_{k,l,m}[n]$ as a uniformly distributed random variable in $[0,2\pi)$, while $\tilde{\theta}_{k,l,m}[n]$ is further assumed to be independent over different coherence intervals in each episode as well as over different UAV-user pairs. We name the above channel model for UAV-ground communications as the \emph{LoS channel with random phase}.

\begin{figure}[t]
  \centering
  \includegraphics[width=8cm]{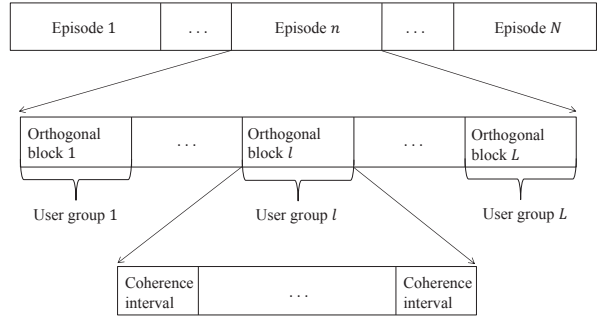}
  \vspace{-3mm}
  \caption{Illustration of the UAV-enabled CoMP transmission and channel model.}\label{Protocol}
  \vspace{-10mm}
\end{figure}

The overall multi-user transmission and channel models over episodes are illustrated in Fig. \ref{Protocol}. Note that each episode $n$ consists of many coherence intervals, and over different coherence intervals the user channels follow the LoS channel with independent random phase but fixed amplitude $\frac{\tau_0}{d_{k,l,m}[n]^2}$'s; however, from episode $n$ to next episode $n+1$, the amplitude values of the LoS channel are changed to $\frac{\tau_0}{d_{k,l,m}[n+1]^2}$'s.
\subsection{ZF Based Beamforming}
At each $n$th episode, when the $l$th group is scheduled for transmission, the received signal at the CP across all the $M$ UAVs is given by\footnote{For the purpose of exposition, this paper assumes that the fronthaul links from all UAVs to the CP are perfect so that the signal distortion induced by the fronthaul transmission is ignored.}
\begin{align}\label{eqn:UAV receive signal}
\mv{y}_l[n]=\sqrt{P}\sum\limits_{k=1}^K\mv{h}_{k,l}[n]s_{k,l}+\mv{z}=\sqrt{P}\mv{H}_l[n]\mv{s}_l+\mv{z},
\end{align}where $s_{k,l}$ denotes the information symbol of the $k$th user of group $l$, which is modeled as independent and identically distributed (i.i.d.) CSCG random variables each with zero mean and unit variance, i.e., $s_{k,l}\sim \mathcal{CN}(0,1),\forall k,l$, and thus $\mv{s}_l=[s_{1,l},\cdots,s_{K,l}]^T$ denotes the transmit symbol vector of group $l$; $P$ denotes the common transmit power of all the users; $\mv{h}_{k,l}[n]=[h_{k,l,1}[n],\cdots,h_{k,l,M}[n]]^T$ denotes the channel vector from the $k$th user of group $l$ to all the UAVs, with $\mv{H}_l[n]=[\mv{h}_{1,l}[n],\cdots,\mv{h}_{K,l}[n]]$; and $\mv{z}=[z_1,\cdots,z_M]^T$ with $z_m\sim \mathcal{CN}(0,\sigma^2)$ denoting the additive white Gaussian noise (AWGN) in the signal forwarded by the $m$th UAV to the CP.

We assume that the channel coefficients between all the users and UAVs, i.e., $\{h_{k,l,m}[n],\forall k,l,m,n\}$, are perfectly known at the CP during each channel coherence interval with the help of channel training and fronthaul transmission. Under this assumption, if the $l$th group is scheduled, the linear beamforming at the CP over the signals from all the UAVs is modeled as
\begin{align}\label{eqn:transmit signal}
\tilde{\mv{s}}_l=\mv{W}_l[n]\mv{y}_l[n]=\sqrt{P}\mv{W}_l[n]\mv{H}_l[n]\mv{s}_l+\mv{W}_l[n]\mv{z},
\end{align}
where $\mv{W}_l[n]=[\mv{w}_{1,l}[n],\cdots,\mv{w}_{K,l}[n]]$ and $\mv{w}_{k,l}[n]$ denotes the beamforming vector for the $k$th user of group $l$ across all UAVs, with $\mv{w}_{k,l}[n]\in \mathbb{C}^{M\times 1}$ and $\|\mv{w}_{k,l}[n]\|=1$.

In this paper, for simplicity we assume that ZF beamforming is adopted to eliminate the inter-user interference for each group of users, i.e., $\mv{w}_{k,l}[n]^H\mv{h}_{j,l}[n]=0$, $\forall k\neq j$. Specifically, define \begin{align}\label{eqn:puesdo inverse}
\bar{\mv{W}}_l[n]=[\bar{\mv{w}}_{1,l}[n],\cdots,\bar{\mv{w}}_{K,l}[n]]=(\mv{H}_l[n]^H\mv{H}_l[n])^{-1}\mv{H}_l[n]^H.
\end{align}
Then, the ZF beamforming vector for the $k$th user of group $l$ at each coherence interval of the $n$th episode is given by
\begin{align}\label{eqn:ZF}
\mv{w}_{k,l}[n]=\frac{\bar{\mv{w}}_{k,l}[n]}{\|\bar{\mv{w}}_{k,l}[n]\|},\quad \forall k,l,n.
\end{align}

With the ZF beamforming solution given in (\ref{eqn:ZF}), the SNR of the $k$th user of group $l$ at each coherence interval of the $n$th episode is given by
\begin{align}
\gamma_{k,l}[n]\ & =\frac{P|\mv{w}_{k,l}[n]^H\mv{h}_{k,l}[n]|^2}{\sigma^2} \label{eqn:SINR 1}
\\ & =\frac{P}{[(\mv{H}_l[n]^H\mv{H}_l[n])^{-1}]_{k,k}\sigma^2},\quad \forall k,l,n. \label{eqn:SINR 2}
\end{align}
The ergodic rate of the $k$th user of group $l$ over the $n$th episode is thus given by
\begin{align}
R_{k,l}[n]&=\frac{1}{L}\mathbb{E}\left[\log_2\left(1+ \frac{P|\mv{w}_{k,l}[n]^H\mv{h}_{k,l}[n]|^2}{\sigma^2}\right)\right] \label{eqn:common rate} \\ & =\frac{1}{L}\mathbb{E} \left[\log_2\left(1+ \frac{P}{[(\mv{H}_l[n]^H\mv{H}_l[n])^{-1}]_{k,k}\sigma^2}\right)\right], ~~~ \forall k,l,n, \label{eqn:common rate ZF}
\end{align}
in bits/second/Hertz (bps/Hz), where the expectation is taken over the random channel matrix $\mv{H}_l[n]$ due to random phase variations.

Finally, the average achievable rate for each user over the $N$ episodes is given by
\begin{align}\label{eqn:average transmit rate}
\bar{R}_{k,l}=\frac{1}{N}\sum\limits_{n=1}^NR_{k,l}[n], ~~~ \forall k,l.
\end{align}

\section{Problem Formulation}\label{sec:Problem Formulation}
In this paper, we aim to optimize the nominal locations of the $M$ UAVs over the $N$ episodes, i.e., $\{x_m[n],y_m[n]\}$, $m=1,\cdots,M$, $n=1,\cdots,N$, to maximize the minimum of the average achievable rates of the $\tilde{K}$ users given in (\ref{eqn:average transmit rate}). The optimization problem is formulated as:
\begin{align}
\!\!\!\!\!\!\mathop{\mathrm{(P1)}\ \underset{\{x_m[n],y_m[n]\}}{\max}} & ~ \min\limits_{1\leq k \leq K,1\leq l \leq L} ~ \bar{R}_{k,l}. \label{eqn:problem}\\
\mathrm{s.t.} \ \ \ & \sqrt{(x_m[n+1]-x_m[n])^2+(y_m[n+1]-y_m[n])^2}\leq D_m[n],\forall m, ~ n=1,\cdots,N-1, \label{eqn:distance of uav}
\end{align}
where $D_m[n]$ denotes the maximum displacement of the $m$th UAV over the $n$th episode, which is determined by the UAV speed limit.

\begin{remark}\label{remark1}
In general, the UAV placement/movement problem can be divided into three cases, depending on the mobility of the UAVs: 1) static UAVs with fixed locations once deployed; 2) high-mobility UAVs with locations changing from one episode to another; and 3) semi-dynamic UAVs, where the locations are changed for some episodes but remain static for the others. Note that our studied problem, i.e., problem (P1), includes all the above three cases: if we set $D_m[n]=0$, $\forall m,n$, then the UAVs are static over all the $N$ episodes; if we set $D_m[n]>0$, $\forall m,n$, then the UAVs are flying with higher mobility as $D_m[n]$'s increase; if we set $D_m[n]=0$ for some episodes, and $D_m[n]>0$ for the others, then the UAVs are deployed in a semi-dynamic manner.
\end{remark}

Note that problem (P1) is generally challenging to solve since the users' ergodic rates at each $n$th episode, i.e., $R_{k,l}[n]$'s shown in (\ref{eqn:common rate}) or (\ref{eqn:common rate ZF}), are difficult to be expressed in closed-form without the expectation operation. To overcome this challenge, we provide an effective and tractable characterization of $R_{k,l}[n]$'s in the next section by applying efficient bounding and approximation techniques.

\section{User Ergodic Rate Characterization}\label{sec:User Average Common Rate Characterization}
In this section, we first propose approximated upper and lower bounds for $R_{k,l}[n]$'s by assuming that the user channels follow independent Rayleigh fading within each episode, i.e.,
\begin{align}\label{eqn:Rayleigh fading}
h_{k,l,m}[n] \sim\mathcal{CN}\left(0,{\frac{\tau_0}{d_{k,m}[n]^2}}\right), ~~~ \forall k,l,m,n.
\end{align}Then, we show that the derived upper and lower bounds for the case of Rayleigh fading are also appropriate bounds of $R_{k,l}[n]$'s for our considered LoS channel model with random phase given in (\ref{LoS2}). Finally, we provide one numerical example, which validates that the obtained analytical bounds are very tight for $R_{k,l}[n]$'s under different setups.

\subsection{Rayleigh Fading Channel}
For ease of exposition, we define $\tilde{R}_{k,l}[n]$ as the ergodic user rate at the $n$th episode, i.e., $R_{k,l}[n]$ as given in (\ref{eqn:common rate}) or (\ref{eqn:common rate ZF}), under the Rayleigh fading channel model given in (\ref{eqn:Rayleigh fading}). To characterize $\tilde{R}_{k,l}[n]$, we start by providing its upper bound and lower bound.
\begin{lemma}\label{lemma1}
At each episode $n$, $\tilde{R}_{k,l}[n]$ is bounded by\begin{align}\label{eqn:convex concave}
R_{k,l}^{\rm lower}[n] \leq \tilde{R}_{k,l}[n] \leq R_{k,l}^{\rm upper}[n], ~~~ \forall k,l,n,
\end{align}
where
\begin{align}
& R_{k,l}^{\rm lower}[n]= \log_2\left(1+ \frac{P}{ \mathbb{E}\left[[(\mv{H}_l[n]^H_l\mv{H}[n])^{-1}]_{k,k}\right]\sigma^2}\right), \label{eqn:common rate lower bound} \\
& R_{k,l}^{\rm upper}[n]= \log_2\left(1+ \frac{P\mathbb{E}\left[\left|\mv{w}_{k,l}[n]^H\mv{h}_{k,l}[n]\right|^2\right]}{\sigma^2}\right). \label{eqn:common rate upper bound}
\end{align}
\end{lemma}

\begin{IEEEproof}
Please refer to Appendix \ref{appendix1}.
\end{IEEEproof}

To obtain the above upper and lower bounds for $\tilde{R}_{k,l}[n]$, we need to derive $\mathbb{E}\left[[(\mv{H}[n]_l^H\mv{H}_l[n])^{-1}]_{k,k}\right]$ and $\mathbb{E}\left[\left|\mv{w}_{k,l}[n]^H\mv{h}_{k,l}[n]\right|^2\right]$ with ZF-based beamforming, respectively. However, as indicated in \cite{Heath11} and \cite{Wei14}, one main difficulty for deriving these expectations lies in that the elements of $\mv{h}_{k,l}[n]$'s are not identically distributed, since the path loss from the $k$th user of group $l$ to different UAVs are in general different. Following the standard technique used in \cite{Heath11} and \cite{Wei14}, we consider the following assumption in the sequel.

\begin{assumption}\label{assumption1}
With the Rayleigh fading channel model in (\ref{eqn:Rayleigh fading}), the channel from the $k$th user of group $l$ to all the UAVs at the $n$th episode, i.e., $\mv{h}_{k,l}[n]$, can be treated as an isotropic vector distributed according to $\mathcal{CN}\left(\mv{0},\frac{\tau_0\sum\limits_{m=1}^M d_{k,l,m}[n]^{-2}}{M}\mv{I}\right)$.
\end{assumption}

Under the above assumption that the total channel power is isotropically distributed in all the directions, we have the following theorems to characterize the upper bound and lower bound of $\tilde{R}_{k,l}[n]$ given in (\ref{eqn:convex concave}).

\begin{theorem}\label{theorem1}
Under Assumption \ref{assumption1}, we have
\begin{align}\label{eqn:upper bound}
R_{k,l}^{\rm upper}[n]=\log_2\left(1+\frac{P\tau_0\sum\limits_{m=1}^Md_{k,l,m}[n]^{-2}}{\frac{M\sigma^2}{M-K+1}}\right), ~~~ \forall k,l,n.
\end{align}
\end{theorem}

\begin{IEEEproof}
Please refer to Appendix \ref{appendix2}.
\end{IEEEproof}

\begin{theorem}\label{theorem2}
Under Assumption \ref{assumption1}, we have
\begin{align}\label{eqn:lower bound}
R_{k,l}^{\rm lower}[n]=\log_2\left(1+\frac{P\tau_0\sum\limits_{m=1}^Md_{k,l,m}[n]^{-2}}{\frac{M\sigma^2}{M-K}}\right), ~~~ \forall k,l,n.
\end{align}
\end{theorem}

\begin{IEEEproof}
Please refer to Appendix \ref{appendix3}.
\end{IEEEproof}

Interestingly, it is observed from (\ref{eqn:upper bound}) in Theorem \ref{theorem1} and (\ref{eqn:lower bound}) in Theorem \ref{theorem2} that the upper bound and lower bound of the user ergodic rate under Rayleigh fading channel model are very close to each other, while the only difference lies in $\frac{M}{M-K+1}$ versus $\frac{M}{M-K}$ in the denominators. As a result, it is expected that both the upper bound (\ref{eqn:upper bound}) and lower bound (\ref{eqn:lower bound}) are very tight for $\tilde{R}_{k,l}[n]$, especially when $M\gg K$.

\subsection{LoS Channel with Random Phase}\label{sec:Approximated Upper Bound and Lower Bound}
It is observed that the only difference between our considered channel model (\ref{LoS2}) and the Rayleigh fading channel (\ref{eqn:Rayleigh fading}) lies in the amplitude: our LoS-based channel model has a fixed amplitude but the Rayleigh fading channel model has a random amplitude. In this subsection, we show that this difference has little effect on the user ergodic rate, i.e., $R_{k,l}[n]\approx \tilde{R}_{k,l}[n]$. As a result, the ergodic user rate upper bound and lower bound given in (\ref{eqn:upper bound}) and (\ref{eqn:lower bound}) based on the Rayleigh fading channel are good approximations for $R_{k,l}[n]$ under our considered LoS channel model with random phase given in (\ref{LoS2}).

The justification is as follows. Based on the ZF orthogonality property, the beamforming vector $\mv{w}_{k,l}[n]$ for the $k$th user of group $l$ is orthogonal to the subspace spanned by the channel vectors from the other $K-1$ users of the same group to all UAVs, i.e.,
\begin{align}\label{eqn:orthogonality}
\mv{w}_{k,l}[n]\bot {\rm span}(\mv{h}_{1,l}[n],\cdots,\mv{h}_{k-1,l}[n],\mv{h}_{k+1,l}[n],\cdots,\mv{h}_{K,l}[n]), ~~~ \forall k,l,n.
\end{align}As a result, $|\mv{w}_{k,l}[n]^H\mv{h}_{k,l}[n]|^2$ in (\ref{eqn:common rate}) is the power of an $M$-dimensional random vector projected onto an $(M-{\rm rank}(\mv{H}_l[n])+1)$-dimensional beamforming space \cite{Maehara90}. With Rayleigh fading, it is well-known that in the case of $K<M$, all the $K$ columns of $\mv{H}_l[n]$ are linearly independent with each other with probability one, i.e., ${\rm rank}(\mv{H}_l[n])=K$. Similarly, we can extend the above result to the channel model given in (\ref{LoS2}). According to \cite{Tao12}, if $K$ vectors are chosen independently with regards to any distribution on $\mathbb{C}^M$ ($K\leq M$) such that the probability of the vector lying in any particular hyperplane through the origin is zero, then the $K$ vectors are linearly independent with each other with probability one. It can be easily shown that the LoS channel with random phase given in (\ref{LoS2}) satisfies the above condition. As a result, we almost surely have ${\rm rank}(\mv{H}_l[n])=K$. In other words, similar to the Rayleigh fading model, for our considered channel in (\ref{LoS2}), $|\mv{w}_{k,l}[n]^H\mv{h}_{k,l}[n]|^2$ is also the power of an $M$-dimensional random vector projected onto an $(M-K+1)$-dimensional beamforming space. Although the random vector has different distributions in the above two different channel models, the projected random space is the same. It is thus expected that the distributions of $|\mv{w}_{k,l}[n]^H\mv{h}_{k,l}[n]|^2$ are close to each other for the cases of Rayleigh fading and our considered channel in (\ref{LoS2}), and hence $R_{k,l}[n]\approx \tilde{R}_{k,l}[n]$, $\forall k,l,n$.

To summarize, for the LoS channel with random phase given in (\ref{LoS2}), we can still use the upper bound (\ref{eqn:upper bound}) and lower bound (\ref{eqn:lower bound}) derived under the independent Rayleigh fading channel model given in (\ref{eqn:Rayleigh fading}) as good approximated bounds to $R_{k,l}[n]$'s. In the following, we provide one numerical example to verify the above approximations.

\subsection{Numerical Example}
\begin{figure}
\begin{center}
\scalebox{0.6}{\includegraphics*{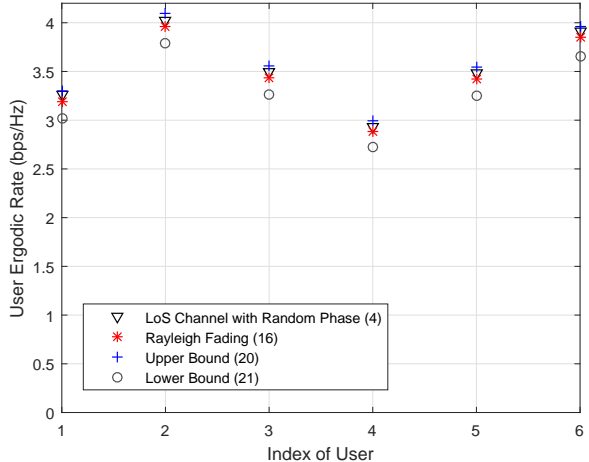}}
\end{center}
\vspace{-5mm}
\caption{Simulated user ergodic rate under the LoS channel with random phase versus various approximations.}\label{fig2}
  \vspace{-10mm}
\end{figure}

In this subsection, we show that the upper bound given in (\ref{eqn:upper bound}) and lower bound given in (\ref{eqn:lower bound}) are tight approximations of $R_{k,l}[n]$'s numerically. In this numerical example, we assume that there are $M=10$ UAVs and $K=6$ users that form one group in the network. Moreover, we just consider one episode with $N=1$, and at this episode, the UAVs and users are randomly located in a $100$ m $\times$ $100$ m square. Moreover, the identical height of UAVs is set as $H=100$ m. The transmit power for each user is $23$ dBm. The power of AWGN at each UAV is $-169$ dBm/Hz, while the channel bandwidth is $10$ MHz. The power of the channel at the reference distance $1$ m is $\tau_0=-40$ dBm. Under this setup, the user ergodic rates under the LoS channel with random phase given in (\ref{LoS2}), under the Rayleigh fading channel given in (\ref{eqn:Rayleigh fading}), and the upper bound and lower bound given in Theorems \ref{theorem1} and \ref{theorem2}, are shown in Fig. \ref{fig2}.

It is observed from Fig. \ref{fig2} that with the Rayleigh fading channel, each user's rate upper bound given in (\ref{eqn:upper bound}) and lower bound given in (\ref{eqn:lower bound}) under Assumption \ref{assumption1} are indeed very tight for $\tilde{R}_{k,l}[n]$'s. This verifies the validness of Assumption \ref{assumption1}. Moreover, it is observed that $\tilde{R}_{k,l}[n]$'s achieved under the Rayleigh fading channel model is very close to $R_{k,l}[n]$'s achieved under our considered channel model, which is in accordance with our discussions in Section \ref{sec:Approximated Upper Bound and Lower Bound}. As a result, we are ready to user either (\ref{eqn:upper bound}) or (\ref{eqn:lower bound}) as one tight approximation of $R_{k,l}[n]$'s.

\begin{remark}\label{remark2}
For both the user rate upper bound and lower bound shown in (\ref{eqn:upper bound}) and (\ref{eqn:lower bound}), it is observed that each user's rate only depends on its own distance to all the $M$ UAVs. This is because ZF beamforming is applied to null out the inter-user interference, and thus each user's ergodic rate, which depends on the projection of the user's channel vector onto an independent beamforming subspace (see (\ref{eqn:orthogonality})), is not affected by the other users' channels/locations. As a result, under our considered protocol in which the UAVs are quasi-static in each episode to serve users via orthogonal blocks, user grouping, i.e., how to select $K$ users to form each of the $L$ groups, does not affect user rate in our setup with ZF beamforming. This is one advantage of the considered model, since the UAV placement and movement solution to problem (P1) applies to all user groupings.
\end{remark}

\section{Proposed Algorithm for Problem (P1)}\label{sec:Proposed Algorithm}
In Section \ref{sec:User Average Common Rate Characterization}, we provide a pair of approximated upper bound and lower bound for each $R_{k,l}[n]$ as given in (\ref{eqn:upper bound}) and (\ref{eqn:lower bound}), respectively. In this section, we study problem (P1) by replacing $R_{k,l}[n]$'s in its objective function with the approximated lower bound given in (\ref{eqn:lower bound}), for obtaining the worst-case achievable rate. In this case, problem (P1) is approximated by the following problem:
\begin{align}
\!\!\!\!\!\mathrm{(P2)} \underset{\{x_m[n],y_m[n]\}}{\max} & ~ \min\limits_{1\leq k \leq K,1\leq l \leq L} ~ \frac{1}{NL}\sum\limits_{n=1}^N\log_2\left(1+\frac{P\tau_0\sum\limits_{m=1}^Md_{k,l,m}[n]^{-2}}{\frac{M\sigma^2}{M-K}}\right) \label{eqn:problem 1} \\
\mathrm {s.t.} \ \ & (\ref{eqn:distance of uav}), \nonumber
\end{align}
where $d_{k,l,m}[n]$ is a function of the UAV nominal locations as given in (\ref{eqn:distance}).

In the following, we solve problem (P2) in three practical scenarios:
\begin{itemize}
\item \textbf{Dynamic UAV placement with full user location information}. First, we consider the case where the location information of the ground users over the $N$ episodes is known ahead of time such that we can jointly optimize the locations of the $M$ UAVs over all $N$ episodes. This may correspond to the scenario when the user movement is pre-programmed, e.g., moving robots that are executing pre-defined tasks.
\item \textbf{Dynamic UAV placement with current user location information}. Second, we consider the case that only the location information of the ground users at the current episode is known. In this case, we can optimize the UAV locations at each episode separately.
\item \textbf{Static UAV placement}. At last, we study the case when the locations of UAVs are fixed over the $N$ episodes, i.e., $D_m[n]=0$, $\forall m,n$. The obtained result may also be applicable to the conventional RAU/RRH placement optimization problem in DAS/C-RAN with RAUs/RRHs deployed on the ground (with $H=0$), and their locations fixed once deployed.
\end{itemize}

\subsection{Dynamic UAV Placement With Full User Location Information}
\subsubsection{Proposed Solution}
With full information of user locations, we can jointly optimize $x_m[n]$ and $y_m[n]$, $m=1,\cdots,M$, over all the $N$ episodes to maximize the minimum of the user average ergodic rates by solving problem (P2). However, problem (P2) is a non-convex problem since the objective function is non-concave over $x_m[n]$'s and $y_m[n]$'s. By introducing a set of auxiliary variables, we equivalently transform problem (P2) into a more tractable form, as shown below.

\begin{theorem}\label{theorem3}
Problem (P2) is equivalent to the following problem:
\begin{align}
\mathrm{(P2-eqv)}&\nonumber\\
\mathop{\mathrm{max}}_{\{x_m[n],y_m[n],c_{k,l,m}[n],R\}}\ &  R \label{eqn:problem 2} \\
\mathrm {s.t.}\ \ \ \ \ \ \ \ \ & \frac{1}{NL}\sum\limits_{n=1}^N\log_2\left(1+\frac{P\tau_0\sum\limits_{m=1}^Mc_{k,l,m}[n]}{\frac{M\sigma^2}{M-K}}\right)\geq R, ~ \forall k,l, \label{eqn:rate target 2} \\ & (x_m[n]\!-\!a_{k,l}[n])^2+(y_m[n]\!-\!b_{k,l}[n])^2\!+\!H^2 \leq \frac{1}{c_{k,l,m}[n]},\ \forall k,l,m,n, \label{eqn:auxiliary 1} \\ &
(x_m[n+1]\!-\!x_m[n])^2+(y_m[n+1]\!-\!y_m[n])^2\leq D_m[n]^2,\ \forall m,n. \label{eqn:distance of uav 2}
\end{align}
\end{theorem}

\begin{IEEEproof}
Please refer to Appendix \ref{appendix4}.
\end{IEEEproof}

In problem (P2-eqv), $R$ denotes the user minimum rate, while the auxiliary variables $c_{k,l,m}[n]$'s are introduced to replace $d_{k,l,m}[n]^{-2}$'s in user ergodic rate expressions. In the new problem, the objective function (\ref{eqn:problem 2}), the user rate target constraint (\ref{eqn:rate target 2}), and the UAV movement constraint (\ref{eqn:distance of uav 2}), are all convex. However, the auxiliary constraint (\ref{eqn:auxiliary 1}) is non-convex since $\frac{1}{c_{k,l,m}[n]}$ is convex, rather than concave, over $c_{k,l,m}[n]$. As a result, problem (P2-eqv) is still a non-convex problem.

Nevertheless, we can use the successive convex approximation technique to solve problem (P2-eqv) locally optimally. Specifically, we provide one concave lower bound for $\frac{1}{c_{k,l,m}[n]}$ as follows. Since $\frac{1}{c_{k,l,m}[n]}$ is convex over $c_{k,l,m}[n]$, its first-order approximation at any given point $\tilde{c}_{k,l,m}[n]>0$:
\begin{align}\label{eqn:first order approximation}
g(c_{k,l,m}[n],\tilde{c}_{k,l,m}[n])=\frac{1}{\tilde{c}_{k,l,m}[n]}-\frac{1}{ \tilde{c}_{k,l,m}[n]^2}(c_{k,l,m}[n]-\tilde{c}_{k,l,m}[n]),
\end{align}serves as its concave lower bound.

Given any point $\tilde{c}_{k,l,m}[n]>0$'s, with $\frac{1}{c_{k,l,m}[n]}$'s replaced by $g(c_{k,l,m}[n],\tilde{c}_{k,l,m}[n])$'s in problem (P2-eqv), we can solve the following approximated convex problem:
\begin{align}
\!\!\!\!\!\!\!\!\!\mathop{\mathrm{max}}_{\{x_m[n],y_m[n],c_{k,l,m}[n],R\}} &  R \label{eqn:problem 3} \\
\mathrm {s.t.}\ \ \ \ \ \ \ \ & \!\!\!(x_m[n]\!-\!a_{k,l}[n])^2\!+\!(y_m[n]\!-\!b_{k,l}[n])^2\!+\!H^2\! \leq\! g(c_{k,l,m}[n],\tilde{c}_{k,l,m}[n]),\!  \forall k,l,m,n,\!\!\! \label{eqn:auxiliary 3} \\ & \!\!\! (\ref{eqn:rate target 2}), ~ (\ref{eqn:distance of uav 2}), \nonumber
\end{align}
where $g(c_{k,l,m}[n],\tilde{c}_{k,l,m}[n])$'s are given in (\ref{eqn:first order approximation}).
Problem (\ref{eqn:problem 3}) is a convex problem, thus can be solved efficiently by CVX \cite{Grant11}.

After solving problem (\ref{eqn:problem 3}) given any point $\tilde{c}_{k,l,m}[n]>0$, the successive convex approximation method for problem (P2-eqv) proceeds by iteratively updating $\tilde{c}_{k,l,m}[n]$'s based on the solution to problem (\ref{eqn:problem 3}). The proposed iterative algorithm is summarized in Algorithm \ref{table2}, where $q$ denotes the index of iteration, and $\epsilon>0$ is a small value to control the convergence of the algorithm. The convergence of Algorithm \ref{table2} is guaranteed by the following theorem.

\begin{algorithm}[t]
{\bf Initialization}: Set the initial values for $\tilde{c}_{k,l,m}[n]>0$'s and $q=1$; \\
{\bf Repeat}:
\begin{enumerate}
\item Find the optimal solution to problem (\ref{eqn:problem 3}) using CVX as $\{x_m[n]^{(q)},y_m[n]^{(q)},c_{k,l,m}[n]^{(q)},R^{(q)}\}$;
\item Update $\tilde{c}_{k,l,m}[n]=c_{k,l,m}[n]^{(q)}$, $\forall k,l,m,n$;
\item $q=q+1$.
\end{enumerate}
{\bf Until} $R^{(q)}-R^{(q-1)}\leq \epsilon$.
\caption{Proposed Algorithm for Solving Problem (P2-eqv).}
\label{table2}
\end{algorithm}

\begin{theorem}\label{theorem4}
Monotonic convergence of Algorithm \ref{table2} is guaranteed, i.e., $R^{(q)}\geq R^{(q-1)}$, $\forall q\geq 2$. Moreover, the converged solution satisfies all the KKT conditions of problem (P2-eqv).
\end{theorem}
\begin{IEEEproof}
The proof of Theorem \ref{theorem4} directly follows that for \cite[Theorem 1]{Marks78}, and is thus omitted for brevity.
\end{IEEEproof}

\subsubsection{Insights on Optimal UAV Placement}
After solving problem (P2-eqv) locally optimally, we aim to gain more insights on the obtained UAV locations. Let $\mu_{k,l,n}^\ast$'s, $\beta_{m,n}^\ast\geq 0$'s, and $\lambda_{k,m,n}^\ast\geq 0$'s denote the optimal dual variables associated with constraints (\ref{eqn:rate target 2}), (\ref{eqn:distance of uav 2}), and (\ref{eqn:auxiliary 3}) in problem (\ref{eqn:problem 3}). Then, the optimal UAV locations over different episodes are summarized by the following lemma.

\begin{lemma}\label{lemma2}
For any $m$th UAV, $m=1,\cdots,M$, its optimal locations over x-axis in the $N$ episodes, i.e., $x_m^{\ast}[n]$, $n=1,\cdots,N$, are the solutions to the following $N$ linear equations:
\begin{align}\label{eqn:optimal primary solution a}
x_m^{\ast}[n]=\frac{\sum\limits_{k=1}^K\sum\limits_{l=1}^L\lambda_{k,l,m,n}^\ast a_{k,l}[n]+\beta_{m,n}^\ast x_m^{\ast}[n+1]+\beta_{m,n-1}^\ast x_m^{\ast}[n-1]}{\sum\limits_{k=1}^K\sum\limits_{l=1}^L\lambda_{k,l,m,n}^\ast+\beta_{m,n}^\ast+\beta_{m,n-1}^\ast}, ~ n=1,\cdots,N.
\end{align}
Note that in (\ref{eqn:optimal primary solution a}), it is assumed that $x_m^{\ast}[N+1]=x_m^{\ast}[-1]=\beta_{m,N}^\ast=\beta_{m,-1}^\ast=0$. Similarly, its optimal locations over y-axis in the $N$ episodes, $y_m^{\ast}[n]$, $n=1,\cdots,N$, are the solutions to the following $N$ linear equations:
\begin{align}\label{eqn:optimal primary solution b}
y_m^{\ast}[n]=\frac{\sum\limits_{k=1}^K\sum\limits_{l=1}^L\lambda_{k,l,m,n}^\ast b_{k,l}[n]+\beta_{m,n}^\ast y_m^{\ast}[n+1]+\beta_{m,n-1}^\ast y_m^{\ast}[n-1]}{\sum\limits_{k=1}^K\sum\limits_{l=1}^L\lambda_{k,l,m,n}^\ast+\beta_{m,n}^\ast+\beta_{m,n-1}^\ast}, ~ n=1,\cdots,N,
\end{align}
where it is assumed that $y_m^{\ast}[N+1]=y_m^{\ast}[-1]=0$.
\end{lemma}

\begin{IEEEproof}
Please refer to Appendix \ref{appendix5}.
\end{IEEEproof}

The intuition behind Lemma \ref{lemma2} is as follows. It is observed from (\ref{eqn:optimal primary solution a}) and (\ref{eqn:optimal primary solution b}) that at any episode $n$, the optimal location of UAV $m$ is the weighted average of user locations at the current episode (episode $n$), as well as the locations of UAV $m$ in the next episode (episode $n+1$) and previous episode (episode $n-1$), where the weights are the corresponding optimal dual variables. This is because besides the user locations, each UAV's location also depends on the maximum UAV displacement constraint over two consecutive episodes shown in (\ref{eqn:distance of uav 2}).

\subsection{Dynamic UAV Placement With Current User Location Information}
Next, consider the case that at each episode, only the current user location information is available. In this case, at each episode $n$, we merely optimize the UAV locations $x_m[n]$'s and $y_m[n]$'s based on the user current location information $a_{k,l}[n]$'s and $b_{k,l}[n]$'s to maximize the minimum user ergodic rate $R_{k,l}[n]$, without considering its effect on the user rate in the future. In other words, at episode $n$, we are interested in the following problem:
\begin{align}
\mathrm{(P2-1)}\mathop{\mathrm{max}}_{\{x_m[n],y_m[n],R[n]\}} & \min\limits_{1\leq k\leq K, 1\leq l \leq L} \frac{1}{L}\log_2\left(1+\frac{P\tau_0\sum\limits_{m=1}^Md_{k,l,m}[n]^{-2}}{\frac{M\sigma^2}{M-K}}\right) \label{eqn:problem 5} \\
\mathrm{s.t.} \ \ \ \ \ \ \ \ \ \ &
(x_m[n+1]\!-\!x_m[n])^2+(y_m[n+1]\!-\!y_m[n])^2\leq D_m[n]^2,\ \forall m. \label{eqn:distance of uav 5}
\end{align}
Note that in the above problem, $x_m[n-1]$'s and $y_m[n-1]$'s, $\forall m$, are treated as known information.

Similar to problem (P2-eqv), we can introduce auxiliary variables to transform problem (P2-1) into the following problem:
\begin{align}
\mathrm{(P2-1-eqv)} \\
\mathop{\mathrm{max}}_{\{x_m[n],y_m[n],c_{k,l,m}[n],R[n]\}} & ~ R[n] \label{eqn:problem 4} \\
\mathrm{s.t.} \ \ \ \ \ \ \ \ \ \ \ & \frac{1}{L}\log_2\left(1+\frac{P\tau_0\sum\limits_{m=1}^Mc_{k,l,m}[n]}{\frac{M\sigma^2}{M-K}}\right)\geq R[n], ~ \forall k,l, \label{eqn:rate target 4} \\ & (x_m[n]\!-\!a_{k,l}[n])^2+(y_m[n]\!-\!b_{k,l}[n])^2\!+\!H^2 \leq \frac{1}{c_{k,l,m}[n]},\ \forall k,l,m, \label{eqn:auxiliary 4} \\ &
(x_m[n+1]\!-\!x_m[n])^2+(y_m[n+1]\!-\!y_m[n])^2\leq D_m[n]^2,\ \forall m. \label{eqn:distance of uav 4}
\end{align}
Then, we can use (\ref{eqn:first order approximation}) to approximate $\frac{1}{c_{k,l,m}[n]}$ and apply the successive convex approximation technique to solve problem (P2-1-eqv) locally optimally. Since the algorithm for this case is similar to Algorithm \ref{table2}, we omit it for brevity.

It is worth noting that similar to Lemma \ref{lemma2} for problem (P2-eqv), it can be shown that the optimal UAV locations in this case are in the following forms:
\begin{align}
& x_m^{\ast}[n]=\frac{\sum\limits_{k=1}^K\sum\limits_{l=1}^L\lambda_{k,l,m,n}^\ast a_{k,l}[n]+\beta_{m,n}^\ast x_m^{\ast}[n-1]}{\sum\limits_{k=1}^K\sum\limits_{l=1}^L\lambda_{k,l,m,n}^\ast+\beta_{m,n}^\ast},\quad\forall m,n, \label{eqn:weighted average a} \\
& y_m^{\ast}[n]=\frac{\sum\limits_{k=1}^K\sum\limits_{l=1}^L\lambda_{k,l,m,n}^\ast b_{k,l}[n]+\beta_{m,n}^\ast y_m^{\ast}[n-1]}{\sum\limits_{k=1}^K\sum\limits_{l=1}^L\lambda_{k,l,m,n}^\ast+\beta_{m,n}^\ast},\quad\forall m,n, \label{eqn:weighted average b}
\end{align}
where $\lambda_{k,m,n}^\ast\geq 0$ and $\beta_{m,n}^\ast \geq 0$ are the optimal dual variables associated with constraints (\ref{eqn:auxiliary 4}) and (\ref{eqn:distance of uav 4}) of problem (P2-1-eqv) (with $\frac{1}{c_{k,l,m}[n]}$ replaced by (\ref{eqn:first order approximation})). As a result, the location of each UAV at any episode is the weighted average of the user locations at the current episode as well as its location in the previous episode. It is worth noting that the difference compared to Lemma \ref{lemma2} in the case with full user location information lies in that the UAV locations at any episode only depend on those at the previous episode, rather than at the next episode (due to the lack of user location information in the future).

\subsection{Static UAV Placement}
At last, we consider the case when the UAVs are static. For simplicity, we assume that full user location information is available. In this case, we can still optimize the UAV locations based on Algorithm \ref{table2} by setting $D_m[n]=0$, $\forall m,n$, in problem (P2-eqv). Nevertheless, since UAVs are static, we can define $x_m=x_m[n]$ and $y_m=y_m[n]$, $\forall m,n$. By removing constraint (\ref{eqn:distance of uav 2}) in problem (P2-eqv), we formulate the following optimization problem:
\begin{align}
\!\!\!\!\!\mathrm{(P2-2)}\mathop{\mathrm{max}}_{\{x_m,y_m,c_{k,l,m}[n],R\}} & R \label{eqn:problem 6} \\
\mathrm {s.t.}\ \ \ \ \ \ \ & \frac{1}{NL}\sum\limits_{n=1}^N\log_2\left(1+\frac{P\tau_0\sum\limits_{m=1}^Mc_{k,l,m}[n]}{\frac{M\sigma^2}{M-K}}\right)\geq R, ~ \forall k,l,n, \label{eqn:rate target 6} \\ & (x_m\!-\!a_{k,l}[n])^2+(y_m\!-\!b_{k,l}[n])^2\!+\!H^2 \leq \frac{1}{c_{k,l,m}[n]}, ~ \forall k,l,m,n. \label{eqn:auxiliary 6}
\end{align}
As compared with problem (P2-eqv), the number of variables and constraints in problem (P2-2) is significantly reduced, and thus the complexity for solving problem (P2-2) is much lower. Specifically, we can use (\ref{eqn:first order approximation}) to approximate $\frac{1}{c_{k,l,m}[n]}$ and apply the successive convex approximation technique to solve problem (P2-2) in a similar manner as that for solving problem (P2-eqv).

It is worth noting that similar to Lemma \ref{lemma2}, the optimal location of each $m$th UAV can be shown in the following form for this case:
\begin{align}
& x_m^{\ast}=\frac{\sum\limits_{k=1}^K\sum\limits_{l=1}^L\sum\limits_{n=1}^N\lambda_{k,l,m,n}^\ast a_{k,l}[n]}{\sum\limits_{k=1}^K\sum\limits_{l=1}^L\sum\limits_{n=1}^N\lambda_{k,l,m,n}^\ast},\quad \forall m, \\
& y_m^{\ast}=\frac{\sum\limits_{k=1}^K\sum\limits_{l=1}^L\sum\limits_{n=1}^N\lambda_{k,l,m,n}^\ast b_{k,l}[n]}{\sum\limits_{k=1}^K\sum\limits_{l=1}^L\sum\limits_{n=1}^N\lambda_{k,l,m,n}^\ast},\quad \forall m,
\end{align}
where $\lambda_{k,l,m,n}^\ast\geq 0$'s are the optimal dual variables associated with constraint (\ref{eqn:auxiliary 6}) of problem (P2-2) (with $\frac{1}{c_{m,n}}$ replaced by (\ref{eqn:first order approximation})). As a result, for the case of static UAVs with full information of user locations, the obtained location of each UAV is the weighted average of all user locations over all the $N$ episodes.

\section{Numerical Examples}\label{sec:Numerical Examples}
In this section, we provide numerical results to verify the effectiveness of the proposed design to optimize the locations of UAVs. The setup is as follows. It is assumed that there are $M=10$ UAVs in the sky, and $\tilde{K}=18$ users on the ground. Moreover, we consider a time interval with $N=200$ episodes, and the time duration of each episode is $0.2$ second (s). In the first episode, we assume that all the $18$ users are randomly located in a $500$ m $\times$ $500$ m square, and in the remaining $199$ episodes, each user can randomly move at a constant speed of $v_{{\rm user}}=15$ m/s within the above $500$ m $\times$ $500$ m square. For UAVs, it is assumed that their identical altitude is $H=100$ m. For convenience, we assume that all the UAVs fly with an identical speed limit $v_{{\rm uav}}$ over all the episodes, which can vary from $0$ m/s (static UAVs) to $20$ m/s; accordingly, $D_m[n]$ can vary from $0$ m to $4$ m (which is significantly less compared to the UAV altitude $H=100$ m, as assumed) . For the uplink communications, it is assumed that the transmit power of each ground user is $23$ dBm. Further, the power spectral density of the AWGN at the UAVs is $-169$ dBm/Hz over a channel bandwidth of $10$ MHz. At last, the power of the channel at the reference distance $1$ m is $\tau_0=-40$ dBm.

\subsection{$L=3$ Groups Each with $K=6$ Users}
\begin{figure}[t]
  \centering
  \includegraphics[width=8cm]{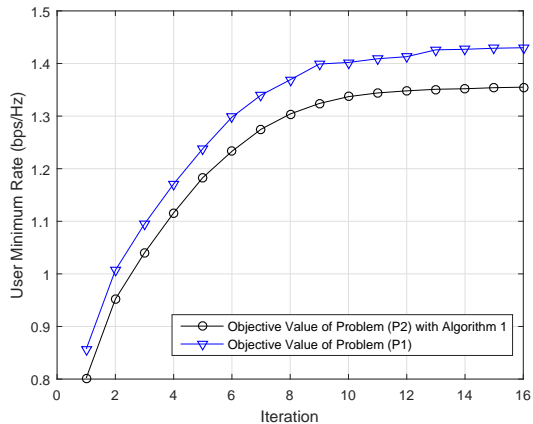}
  \vspace{-3mm}
  \caption{Convergence of Algorithm \ref{table2} with $D_m[n]=20$ m, $\forall m,n$.}\label{Convergence}
  \vspace{-10mm}
\end{figure}

First, consider the case when the ground users are divided to $L=3$ groups, each with $K=6$ users where $K<M=10$ as assumed. Fig. \ref{Convergence} shows the convergence of Algorithm \ref{table2} for solving problem (P2) under the case of dynamic UAV placement with full information of user movement when the maximum speed of UAVs is $v_{{\rm uav}}=10$ m/s, or equivalently, $D_m[n]=2$ m, $\forall m,n$. In this numerical example, in the first iteration of Algorithm \ref{table2}, we randomly generate the locations of UAVs over the $N=200$ episodes (subject to the given constraint on $D_m[n]$'s) as the initial points. Moreover, since problem (P2) is one approximation of problem (P1), we also substitute the UAV location solution obtained by each iteration of Algorithm \ref{table2} back into the objective function of problem (P1) to obtain the actual achievable minimum user rate, as also shown in Fig. \ref{Convergence} for comparison.

From Fig. \ref{Convergence}, a monotonic convergence behaviour is observed for Algorithm \ref{table2}, as expected from Theorem \ref{theorem4}. Moreover, it is observed that problem (P2) is a good approximation (lower bound) to problem (P1), since the objective values of these two problems are very close to each other with the obtained UAV location solution after each iteration. At last, as compared to a random UAV placement strategy (in the first iteration), our optimized UAV location solution after convergence can improve the rate performance by about $70\%$.

\begin{figure}[t]
  \centering
  \includegraphics[width=8cm]{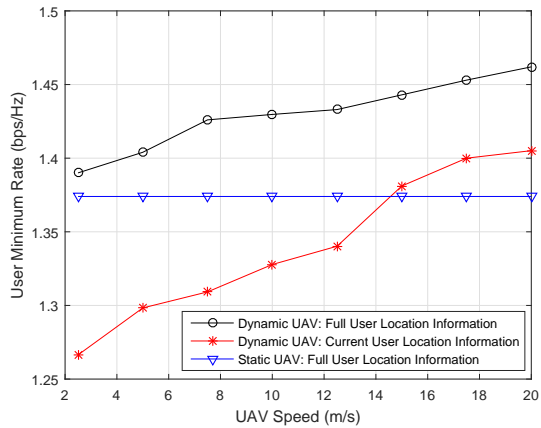}
    \vspace{-3mm}
\caption{User minimum rate versus UAV maximum speed in three scenarios: dynamic UAV with full user location information, dynamic UAV with current user location information, and static UAV with full user location information.}\label{rate}
  \vspace{-10mm}
\end{figure}

Next, we compare the user minimum rate achieved in the three cases of dynamic UAV with full user location information, dynamic UAV with current user location information, and static UAV with full user location information. It is observed from Fig. \ref{rate} that if the full user location information is known, the minimum rate performance is not very sensitive to the maximum UAV speed. Specifically, even with static UAV deployment, the rate loss as compared to the case of dynamic UAV deployment with full user location information and $v_{{\rm uav}}=20$ m/s is about $0.1$ bps/Hz. However, if only current user location is known, user minimum rate increases very fast with UAV speed. This is because without the future user location information, the UAVs should move very fast to maintain a satisfactory service, while with full user location information, we can jointly optimize the UAV locations over all episodes such that the movement of each UAV over two consecutive episodes is usually smaller and thus increasing the UAV maximum speed is less useful. Moreover, it is observed that when $v_{{\rm uav}}\leq 15$ m/s, i.e., users move faster than UAVs, the user minimum rate achieved with partial user location information is much lower than that achieved with full information. However, when UAVs can move faster than users, UAVs have sufficient capability to adjust their locations at each episode to compensate the lack of future user location information, and thus the rate loss is much smaller.

\begin{figure}
\begin{center}
\subfigure[Dynamic UAV with Full User Location Information]{\scalebox{0.5}{\includegraphics*{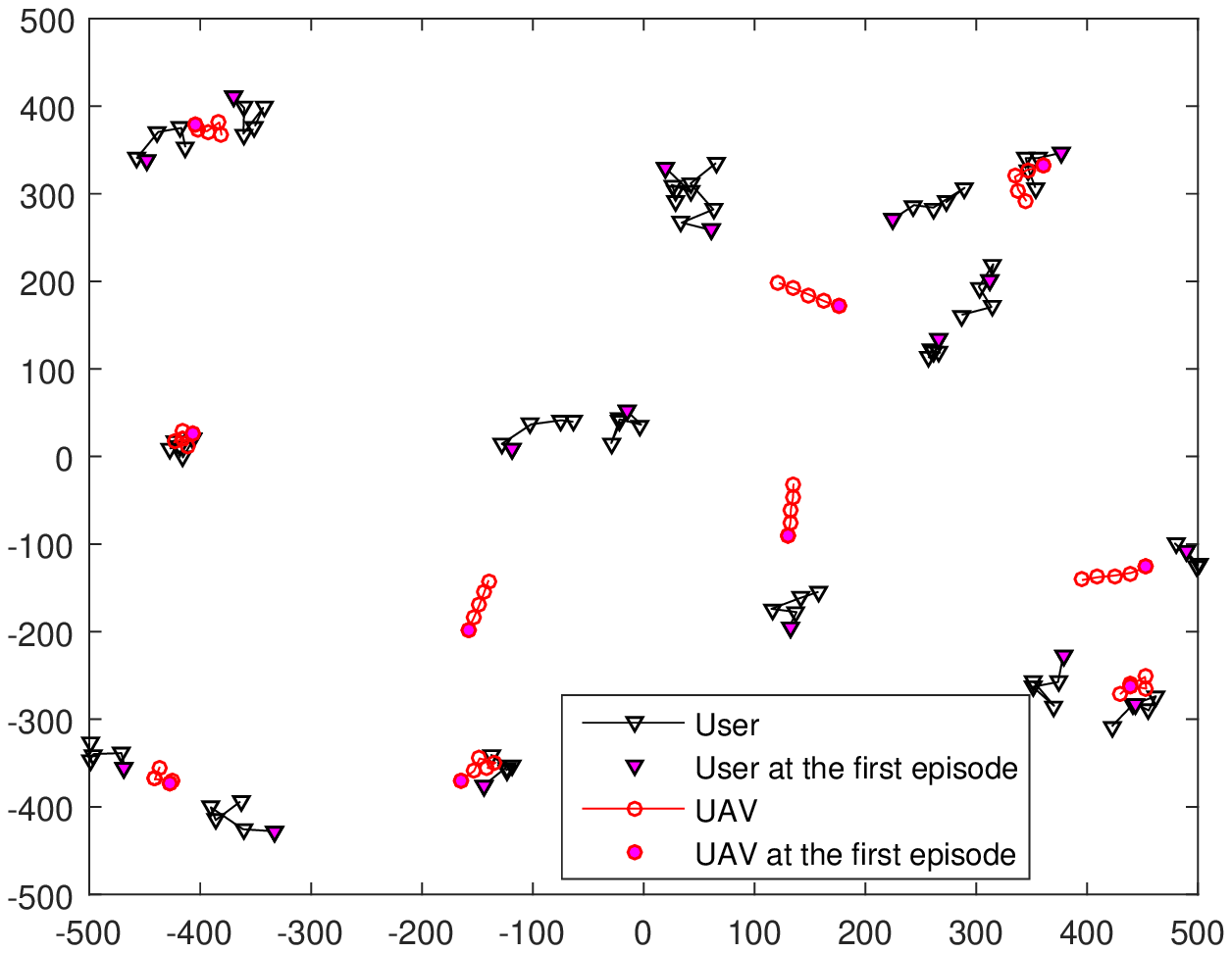}}}
\subfigure[Dynamic UAV with Current User Location Information]{\scalebox{0.5}{\includegraphics*{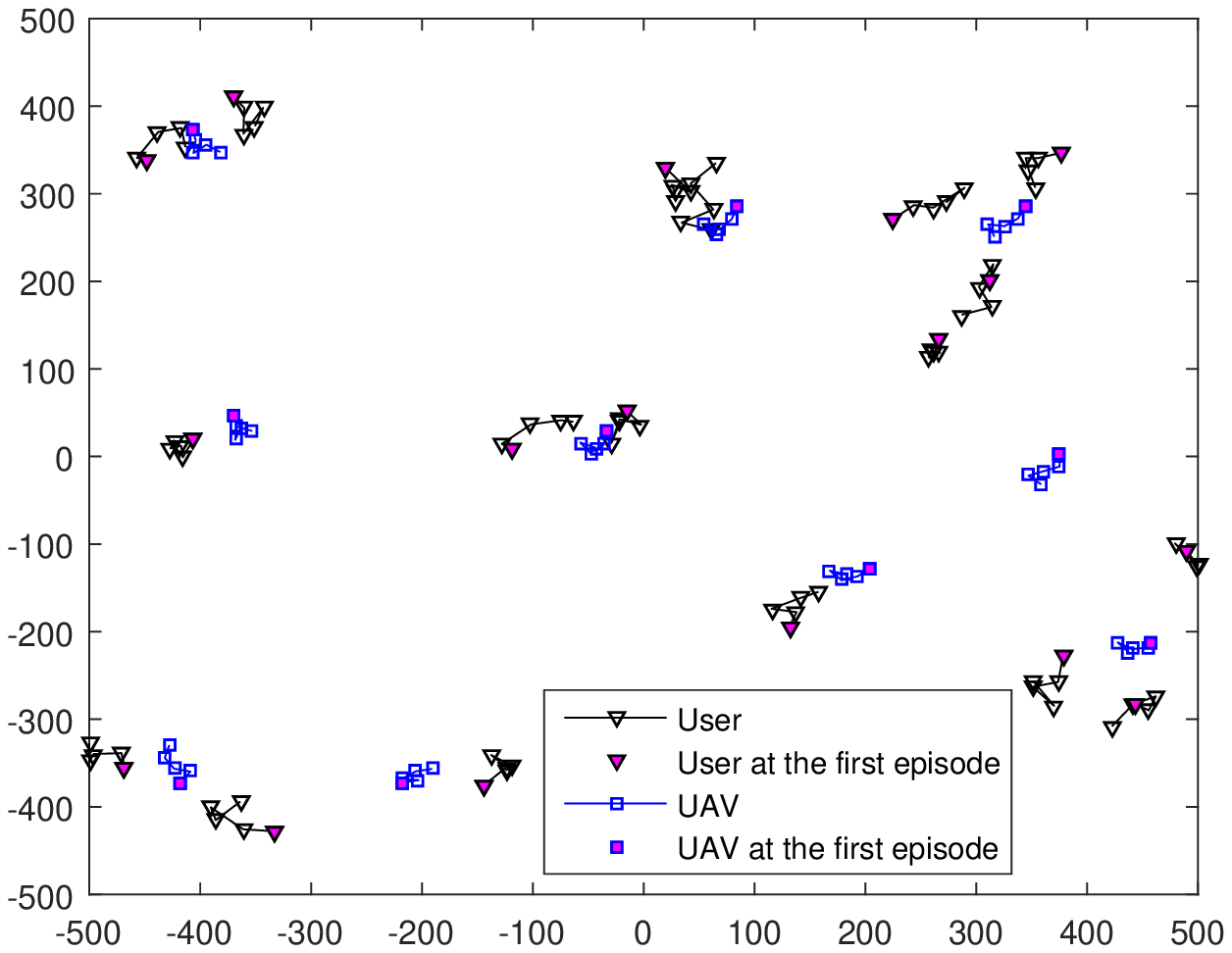}}}
\subfigure[Static UAV with Full User Location Information]{\scalebox{0.5}{\includegraphics*{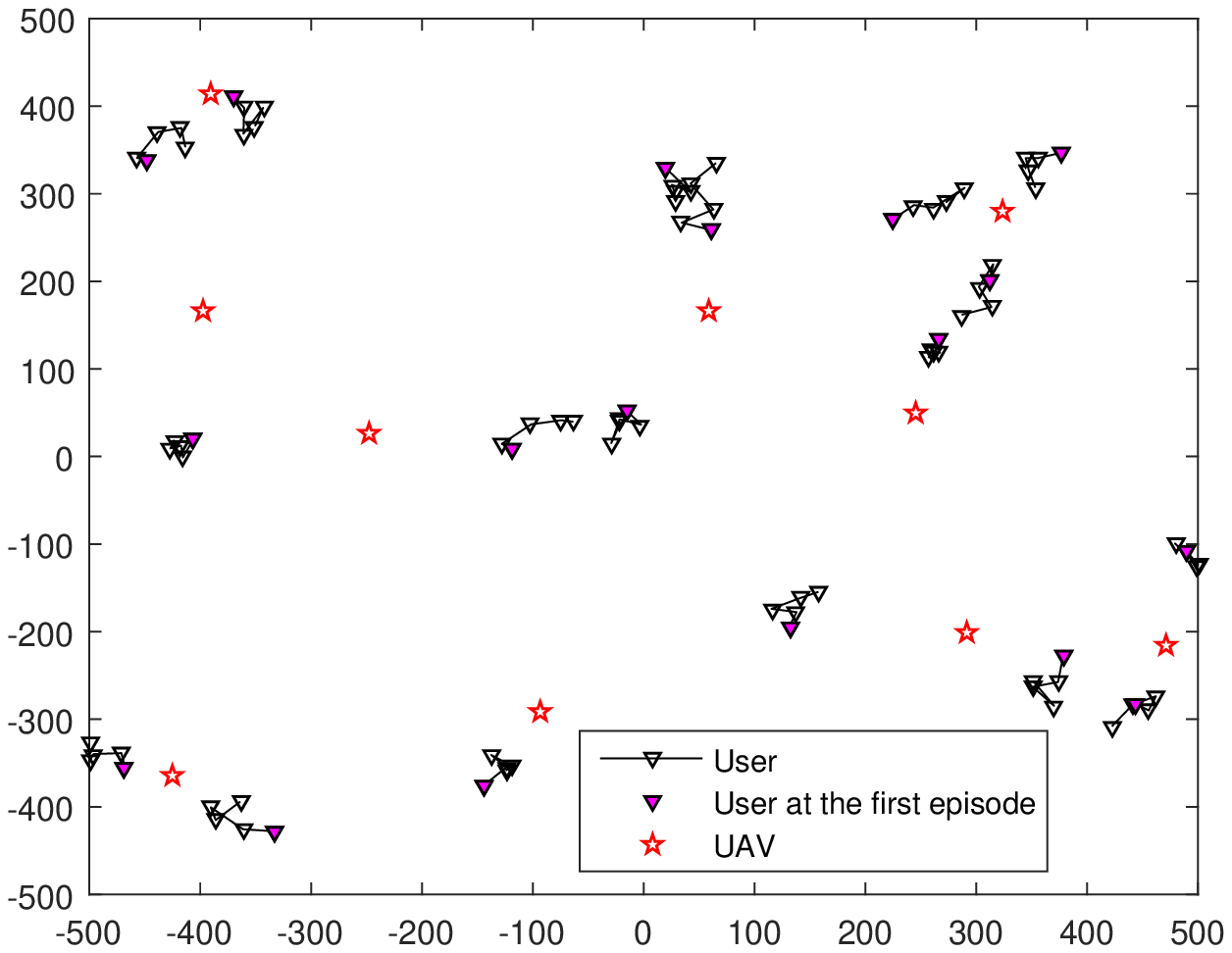}}}
\caption{UAV placement or movement (in the $1$st, $11$st, $21$st, $31$st, $41$st episodes).}\label{movement}
\end{center}
  \vspace{-10mm}
\end{figure}

At last, Fig. \ref{movement} shows the UAV locations versus the user locations in the $1$st, $11$st, $21$st, $31$st, $41$st episodes in the three cases of dynamic UAV with full user location information, dynamic UAV with current user location information, and static UAV with full user location information. It is observed that for all these three cases, if there is any isolated user, then one UAV will be deployed dedicatedly for covering it; while when some users are close to each other, one UAV will be deployed in the middle of them to serve all of them. This verifies that the UAV locations should be the weighted average of the user locations, as shown in Section \ref{sec:Proposed Algorithm}, while the weights for the nearby users are larger. It is also observed that with full user location information, each UAV tends to fly towards one direction, while with current user location information only, due to the lack of a future plan, many UAVs keep flying back and forth.

\subsection{Effect of User Grouping on User Minimum Rate}
According to Remark \ref{remark2}, given any $L$ and $K$ with $LK=\tilde{K}$, how to select $K$ users to form one group does not affect the resulting user achievable rate thanks to the ZF-based beamforming. However, the number of groups (and thus the number of users per group) affects the achievable rate. If there are fewer groups each with more users, on one hand, each group can be scheduled for transmission with more time/bandwidth; while on the other hand, in the scheduled interval, the ZF gain is smaller. As a result, there exists a trade-off in designing the number of groups of the ground users.

\begin{figure}[t]
  \centering
  \includegraphics[width=8cm]{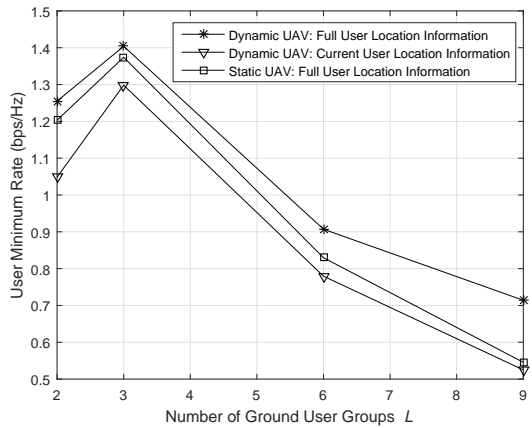}
    \vspace{-3mm}
\caption{User minimum rate comparison for the cases of $L=2,3,6,9$ ($K=9,6,3,2$) when the maximum UAV speed is $v_{{\rm uav}}=5$ m/s.}\label{group}
  \vspace{-10mm}
\end{figure}

Fig. \ref{group} compares the user minimum rate performance in the cases of $L=2,3,6,9$ groups of ground users, or equivalently $K=9,6,3,2$ ground users per group, when the maximum UAV speed is $v_{{\rm uav}}=5$ m/s. It is observed that in this numerical example, the user minimum rate first increases with $L$, and then decreases with $L$, under all the cases of dynamic UAV with full user location information, dynamic UAV with current user location, and static UAV with full user location information. Moreover, dividing the users into $L=3$ groups is optimal in terms of user minimum rate maximization.

%
%

\vspace{-5pt}
\section{Conclusion}\label{sec:Conclusion}
This paper studies the minimum rate maximization problem in the uplink communications of a multi-UAV enabled multi-user network, in which the placement and movement of the UAVs are designed to achieve maximum average throughput of the moving users on the ground. In contrast to the existing literature on UAV-enabled communications, this paper considers the new setup where UAVs serve as flying RAUs or RRHs for relaying the user messages to the CP for joint decoding. Since the performance of CoMP based communication depends on both the amplitude and phases of user channels, this paper proposes a new LoS channel model with random phase for UAV-enabled CoMP, under which very tight approximations of user ergodic rates are characterized in closed-form by means of random matrix theory. With such an analytic ergodic rate expression, we propose an efficient algorithm to design the placement and dynamic movement of UAVs based on the successive convex approximation technique. Numerical results are provided to show the effectiveness of the proposed algorithm under various practical application scenarios.

\vspace{-5pt}
\begin{appendix}
\subsection{Proof of Lemma \ref{lemma1}}\label{appendix1}
It can be shown that given $a,b>0$, $f(x)=\log_2(1+ax)$ is concave over $x>0$, and $g(x)=\log_2(1+b/x)$ is convex over $x>0$. As a result, according to Jensen's inequality to (\ref{eqn:common rate}) and (\ref{eqn:common rate ZF}), $R_{k,l}^{\rm upper}[n]$ given in (\ref{eqn:upper bound}) and $R_{k,l}^{\rm lower}[n]$ given in (\ref{eqn:lower bound}) serve as the upper bound and lower bound of $\tilde{R}_{k,l}[n]$, respectively. Lemma \ref{lemma1} is thus proved.

\subsection{Proof of Theorem \ref{theorem1}}\label{appendix2}
Based on the orthogonal property of ZF beamforming, the beamforming vector $\mv{w}_{k,l}[n]$ for the $k$th user of group $l$ is orthogonal to the subspace spanned by the channel vectors from the other $K-1$ users in the same group to all the UAVs, i.e.,
\begin{align}
\mv{w}_{k,l}[n] \bot {\rm span}(\mv{h}_{1,l}[n],\cdots,\mv{h}_{k-1,l}[n],\mv{h}_{k+1,l}[n],\cdots,\mv{h}_{K,l}[n]), ~~~ \forall k,l,n.
\end{align}
It is well-known that $\mv{h}_{k,l}[n]$'s are linearly independent with each other with probability one if the elements in each $\mv{h}_{k,l}[n]$ are i.i.d. Gaussian random variables. As a result, under Assumption \ref{assumption1}, $|\mv{w}_{k,l}[n]^H\mv{h}_{k,l}[n]|^2$ is the power of an isotropic $M$-dimensional random vector projected onto an $(M-K+1)$-dimensional beamforming space \cite{Marks78}. According to \cite[Theorem 1.1]{Marks78}, we have
\begin{align}
|\mv{w}_{k,l}[n]^H\mv{h}_{k,l}[n]|^2\sim \Gamma\left(\frac{M-K+1}{M},\sum\limits_{m=1}^Md_{k,l,m}[n]^{-2}\right), ~~~ \forall k,l,n.
\end{align}
Consequently, we have
\begin{align}\label{eqn:expectation signal}
\mathbb{E}[|\mv{w}_{k,l}[n]^H\mv{h}_{k,l}[n]|^2]=(M-K+1)\sum\limits_{m=1}^Md_{k,l,m}[n]^{-2}/M,\quad \forall k,l,n.
\end{align}
By substituting (\ref{eqn:expectation signal}) into (\ref{eqn:common rate upper bound}), Theorem \ref{theorem1} is thus proved.

\subsection{Proof of Theorem \ref{theorem2}}\label{appendix3}
Under Assumption \ref{assumption1}, we can model the UAV-user channel matrix as
$\mv{H}_l[n]=\mv{G}_l[n]\mv{\Phi}_l[n]^{\frac{1}{2}}$,
where $\mv{G}_l[n]\sim \mathcal{CN}(\mv{0},\mv{I})$, and $\mv{\Phi}_l[n]$ is a diagonal matrix with the $k$th diagonal element denoted by $[\mv{\Phi}_l[n]]_{k,k}=\sum\limits_{m=1}^Md_{k,l,m}[n]^{-2}/M$, $\forall k,l,n$. As a result, we have
\begin{align}
\mathbb{E}[[(\mv{H}_l[n]^H\mv{H}_l[n])^{-1}]_{k,k}] & = \mathbb{E}[[(\mv{\Phi}_l[n]^{\frac{1}{2}}\mv{G}_l[n]^H\mv{G}_l[n]\mv{\Phi}_l[n]^{\frac{1}{2}})^{-1}]_{k,k}] \\
& = \mathbb{E}[[\mv{\Phi}_l[n]^{-\frac{1}{2}}(\mv{G}_l[n]^H\mv{G}_l[n])^{-1}\mv{\Phi}_l[n]^{-\frac{1}{2}}]_{k,k}] \\
& = \frac{1}{\mv{\Phi}_l[n]_{k,k}}\mathbb{E}[[(\mv{G}_l[n]^H\mv{G}_l[n])^{-1}]_{k,k}] \\
& = \frac{M}{\sum\limits_{m=1}^Md_{k,l,m}[n]^{-2}}\mathbb{E}[[(\mv{G}_l[n]^H\mv{G}_l[n])^{-1}]_{k,k}].
\end{align}
Note that since $\mv{G}_l[n]\sim \mathcal{CN}(\mv{0},\mv{I})$, $(\mv{G}_l[n]^H\mv{G}_l[n])^{-1}$ is a Wishart Matrix \cite{Verdu04}. Since $\mv{G}_l[n]\in \mathbb{C}^{M\times K}$, according to \cite[Lemma 2.10]{Verdu04}, we have
\begin{align}
\mathbb{E}[{\rm tr}((\mv{G}_l[n]^H\mv{G}_l[n])^{-1})]=\frac{K}{M-K}.
\end{align}
It thus follows that
\begin{align}
&\mathbb{E}[[(\mv{H}_l[n]^H\mv{H}_l[n])^{-1}]_{k,k}]=\frac{M}{\sum\limits_{m=1}^Md_{k,l,m}[n]^{-2}}\mathbb{E}[[(\mv{G}_l[n]^H\mv{G}_l[n])^{-1}]_{k,k}] \\
=& \frac{M}{\sum\limits_{m=1}^Md_{k,l,m}[n]^{-2}} \frac{\mathbb{E}[[{\rm tr}((\mv{G}_l[n]^H\mv{G}_l[n])^{-1})]}{K}= \frac{M}{\sum\limits_{m=1}^Md_{k,l,m}[n]^{-2}} \frac{K}{(M-K)K} \\
=& \frac{M}{(M-K)\sum\limits_{m=1}^Md_{k,l,m}[n]^{-2}}.
\end{align}
By substituting the above result into (\ref{eqn:common rate lower bound}), Theorem \ref{theorem2} is thus proved.

\subsection{Proof of Theorem \ref{theorem3}}\label{appendix4}
It can be shown that by introducing a minimum rate variable $R$, problem (P2) is equivalent to the following problem:
\begin{align}
\!\!\!\!\!\!\underset{\{x_m[n],y_m[n],R\}}{\max} & ~ R ~ \label{eqn:problem rate} \\
\mathrm {s.t.} \ \ & ~ \frac{1}{NL}\sum\limits_{n=1}^N\log_2\left(1+\frac{P\tau_0\sum_{m=1}^Md_{k,l,m}[n]^{-2}}{M\sigma^2/(M-K)}\right)\geq R, ~ \forall k,l,n, \label{eqn:rate teraget} \\ & ~ (\ref{eqn:distance of uav}). \nonumber
\end{align}

First, given any solution $x_m[n]$'s and $y_m[n]$'s to problem (\ref{eqn:problem rate}), we can set $x_m[n]$'s and $y_m[n]$'s and \begin{align}\label{eqn:feasible}
c_{k,l,m}[n]=\frac{1}{(x_m[n]-a_{k,l}[n])^2+(y_m[n]-b_{k,l}[n])^2+H^2}, ~~~ \forall k,l,m,n,
\end{align}
as one feasible solution to problem (P2-eqv), which achieves the same objective value. As a result, the optimal value of problem (P2-eqv) is no smaller than that of problem (\ref{eqn:problem rate}).

Moreover, since the objective value of problem (P2-eqv) is an increasing function of $c_{k,l,m}[n]$'s, with the optimal solution, (\ref{eqn:feasible}) must be true. As a result, the optimal value of problem (P2-eqv) is achievable by problem (P2). In other words, the optimal value of problem (P2-eqv) is no larger than that of problem (\ref{eqn:problem rate}).

To summarize, the optimal value of problem (\ref{eqn:problem rate}) is the same as that of problem (P2-eqv). Theorem \ref{theorem3} is thus proved.

\subsection{Proof of Lemma \ref{lemma2}}\label{appendix5}
The Lagrangian of problem (\ref{eqn:problem 3}) is:
\begin{align}\label{eqn:Lagrange}
&\mathcal{L}(\{x_m[n],y_m[n],c_{k,l,m}[n],\mu_{k,l,n},\lambda_{k,l,m,n},\beta_{m,n}\}) \nonumber \\ =&R+\sum\limits_{k=1}^K\sum\limits_{l=1}^L\sum\limits_{n=1}^N\mu_{k,l,n}\left(\frac{1}{NL}\sum\limits_{n=1}^N\log_2\left(1+\frac{P\tau_0\sum\limits_{m=1}^Mc_{k,l,m}[n]}{\frac{M\sigma^2}{M-K}}\right)-R\right) \nonumber \\ & -\sum\limits_{m=1}^M\sum\limits_{k=1}^K\sum\limits_{l=1}^L\sum\limits_{n=1}^N\lambda_{k,l,m,n}\bigg((x_m[n]-a_{k,l}[n])^2\nonumber \\ & +(y_m[n]-b_{k,l}[n])^2+H^2-\frac{1}{\tilde{c}_{k,l,m}[n]}+\frac{1}{ \tilde{c}_{k,l,m}[n]^2}(c_{k,l,m}[n]-\tilde{c}_{k,l,m}[n])\bigg) \nonumber \\
& -\sum\limits_{m=1}^M\sum\limits_{n=1}^{N-1}\beta_{m,n}\bigg((x_m[n+1]-x_m[n])^2+(y_m[n+1]-y_m[n])^2-D_m[n]^2\bigg),
\end{align}
where $\mu_{k,l,n}$'s, $\beta_{m,n}\geq 0$'s, and $\lambda_{k,l,m,n}\geq 0$'s are the dual variables associated with constraints (\ref{eqn:rate target 2}), (\ref{eqn:distance of uav 2}), and (\ref{eqn:auxiliary 3}) in problem (\ref{eqn:problem 3}), respectively.

The dual function of problem (\ref{eqn:problem 3}) is then given by
\begin{align}\label{eqn:dual function}
g(\{\mu_{k,l,n},\lambda_{k,l,m,n},\beta_{m,n}\})=\max \limits_{\{x_m[n],y_m[n],c_{k,l,m}[n]\}} ~ \mathcal{L}(\{x_m[n],y_m[n],c_{k,l,m}[n],\mu_{k,l,n},\lambda_{k,l,m,n},\beta_{m,n}\}).\!\!\!
\end{align}
The dual problem to problem (\ref{eqn:problem 3}) is thus formulated as
\begin{align}
\mathop{\mathrm{min}}_{\{\mu_{k,l,n},\lambda_{k,l,m,n},\beta_{m,n}\}} & ~ g(\{\mu_{k,l,n},\lambda_{k,l,m,n},\beta_{m,n}\})  \label{eqn:dual problem} \\
\mathrm {s.t.} \ \ \  & \mu_{k,l,n}\geq 0, ~ \forall k,l,n, \label{eqn:dual constraint 3}, \\ &  \lambda_{k,l,m,n}\geq 0, ~ \forall k,l,m,n, \label{eqn:dual constraint 1} \\
& \beta_{m,n}\geq 0, ~ \forall m,n. \label{eqn:dual constraint 2}
\end{align}

Let $\mu_{k,l,n}^\ast$'s, $\beta_{m,n}^\ast\geq 0$'s, and $\lambda_{k,m,n}^\ast\geq 0$'s denote the optimal dual variables to problem (\ref{eqn:dual problem}). Given these optimal dual variables, the derivative of the Lagrangian of problem (\ref{eqn:problem 3}) given in (\ref{eqn:Lagrange}) over $x_m[n]$ is
\begin{align}
&\frac{\partial \mathcal{L}(\{x_m[n],y_m[n],c_{k,l,m}[n],\mu_{k,l,n}^\ast,\lambda_{k,l,m,n}^\ast,\beta_{m,n}^\ast\})}{\partial x_m[n]}\\
=&2\sum\limits_{k=1}^K\sum\limits_{l=1}^L\lambda_{k,l,m,n}^\ast(x_m[n]-a_{k,l}[n])
+2\beta_{m,n}^\ast(x_m[n+1]-x_m[n])+2\beta_{m,n-1}^\ast(x_m[n]-x_m[n-1]).\nonumber
\end{align}
By setting the derivative to zero, it can be shown that the optimal $x_m^\ast[n]$ must satisfy (\ref{eqn:optimal primary solution a}). Similarly, it can be shown that the optimal $y_m^\ast[n]$ must satisfy (\ref{eqn:optimal primary solution b}).

Lemma \ref{lemma2} is thus proved.

\end{appendix}

\end{document}